\documentclass[11pt]{article}
\usepackage[utf8]{inputenc}
\usepackage{listings}
\usepackage{graphicx}
\usepackage{natbib}
\usepackage{wrapfig}
\usepackage{csquotes}
\usepackage{geometry}
\title{\textbf{A black market for upvotes and likes}}
\author{Mihály Héder\\
		MTA SZTAKI\\
		mihaly.heder@sztaki.mta.hu\\
		}
\date{}

\begin{document}

\maketitle

\begin{abstract}

\textbf{Purpose}: This research investigates controversial online marketing techniques that involve buying hundreds or even thousands of upvotes, likes, comments, etc. 

\textbf{Methodology}: Observation and categorization of 7,426 \enquote{campaigns} posted on the crowdsourcing platform microworkers.com over a 365 day (i.e., yearlong) period were conducted. Hypotheses about the mechanics and effectiveness of these campaigns were established and evaluated. 

\textbf{Findings}: The campaigns contained a combined 1,856,316 microtasks, 89.7\% of which were connected to online promotion. Techniques for search engine manipulation, comment--generating in the scale of tens of thousands, online vote manipulation, mass account creation, methods for covering tracks were discovered. The article presents an assessment of the effectiveness of such campaigns as well as various security challenges created by these campaigns. 

\textbf{Research limitations}: The observed campaigns form only a small portion of the overall activity. This is due to invite-only campaigns and the presence of alternative, unobservable platforms. 

\textbf{Practical implications}: The findings of this article could be input for detecting and avoiding such online campaigns. 

\textbf{Social implications}: The findings show that in some conditions tremendous levels of manipulation of an online discourse can be achieved with a limited budget.

\textbf{Originality}: While there is related work on \enquote{follower factories} and \enquote{click/troll farms}, those entities offer complete \enquote{solutions} and their techniques are rather opaque. By investigating a crowdsourcing platform, this research unveils the underlying mechanics and organization of such campaigns. The research is based on a uniquely large number of observations. Small, cheap campaigns, the manipulation of less significant platforms is also included, while the related work tends to focus on mass, politically motivated efforts.

\end{abstract}

\section{Introduction}

This article analyses marketing campaigns that have been executed through the hiring of \emph{freelancers} or \enquote{\emph{microworkers}} to complete short, menial tasks called \emph{microtasks} that usually pay less than one US dollar each, and most often only around ten cents. These tasks include watching, liking, upvoting, and \enquote{+1}-ing items on web platforms featuring social media functions, like Facebook, Twitter, Reddit, and Instagram. A job description showing what such a campaign looks is given below (an actual, observed example):
\begin{lstlisting}[caption=Buying facebook likes, label={facelikes}]
Title: Facebook Like: <REDACTED String>
Payment: USD 0.15
Number of workers accepted: 100

Job description:

WARNING: we manually review almost all of the submitted tasks. Thus, if we find that you have ignored the instructions (i.e., posting on a wrong site, using non-unique or nonsensical content), you will be permanently banned from our system. You must have 50 Facebook friends.

1. Go to <REDACTED URL1>
2. Visit the URL shown at <REDACTED URL1>
3. On that page, you will see a Facebook Like button. Click on it
4. Submit your Facebook profile URL on <REDACTED URL1> to verify that you have completed the task (make sure you have set your Facebook profile to Public View in order for your task to be verified) to get a 7 character confirmation code
(...)
\end{lstlisting}

Completing this task pays USD 0.15 for the microworkers. In this case, one hundred freelancers were sought to do the task.

There are specialized web platforms for the brokering of small tasks. One of them is microworkers.com, which was investigated for this article. This platform lets employers run campaigns for a fee, usually 10\% of the payment made to the freelancing internet users who do the job, called microworkers. Therefore, the 100 Likes above would cost the client posting the job only USD 16.5. The platform has hundreds of thousands of microworkers \citep{Nguyen2014}.

The platform was not specifically created for promotional or marketing purposes, and indeed data processing, survey-based research, and software testing jobs are also posted on the platform. However, as shown below, the majority of tasks can be categorized as being of a promotional nature (See Table \ref{ActBudg}).

\subsection{Article Structure}

The structure of the present article is as follows. After this introduction, the terminology is presented and a number of ethical questions are posed. This is followed by an outline of the related work in this field. The next chapter describes the research aims and details of the observation, as well as the limitations of the research. This is followed by the Results section, which also contains a number of separate subchapters covering various aspects of the investigation into the most important gray promotional activities, including estimates about their efficiency, security concerns, and possible counter-measures, where applicable. The Results section also describes some observed campaign-supplementing techniques. Finally, the Conclusions section offers a summary and outlines some possibilities for future work.

\section{Why call it a black market?}

In the opinion of the author of this paper buying likes, followers, votes, upvotes, retweets, etc. (the generic term for these used in this paper is social media activity) for promotional purposes is an unethical practice, therefore we use the term \enquote{black market} to describe this part of the industry. This is not to say they are necessarily breaking any rules or laws—that question is outside the scope of this paper.

This kind of activity is considered unethical for the following reasons. a) The users of a social platform are misled by a page or post having an artificially inflated number of likes, followers, etc. Normally, it is impossible for users to differentiate between paid activity and genuine activity. The conventional semantic behind a like, upvote, or follow is a statement of approval of the content in question. In other words, the microworkers are paid to lie in the sense that they are paid to pretend to like/endorse/approve content that they most likely have not truly read or watched. b) The microworker's \enquote{employment} is arguably rather exploitative, because of the low payment and the apparent lack of any powers against the clients (see table \ref{Payment}). c) Content creators and social media users who do not employ such practices are clearly disadvantaged. Finally, d) it is easy to see that if paid social media activity were more universal, it would create an unsustainable social media environment.

Besides paid social activity, there are other highly controversial campaigns based around social media. One such typical activity involves creating accounts on behalf of a client and then the microworker handing over the user name and password to the client. An example of this kind of activity is given in the following (actual, observed example):

\begin{lstlisting}[caption=Acquiring gmail accounts, label=gmailacc]
Title: Gmail: Create an Account
Payment: 0.16
Number of workers accepted: 230

Job description:
Note: I will Check American Name and Profile Picture otherwise I have to decline you.

1. Go to www.fakenamegenerator.com
2. Choose proper American name
3. Go to Gmail.com
4. Create a new Gmail account using details from www.fakenamegenerator.com
5. Upload a good profile picture in Gmail

For proof:
- Give Password
- Give a Recovery Mail so that I can change it later

Note: Make sure all are perfect otherwise, I will decline your payment.

Required proof that task was finished?
1. Gmail and Username
2. Password
3. Recovery email

\end{lstlisting}

In this case, the client was able to acquire 230 Gmail accounts for a mere \$ 40.48. We might speculate that these accounts (and similarly those created for YouTube, Twitter, etc.) will be used for promotional purposes, controlled by the client, while also raising all the ethical concerns highlighted in the previous example. However, this speculation might actually be optimistic. Fake accounts like these could also be used for more sinister purposes, such as as tools for spreading fake news \citep{Allcott2017} for political purposes or for committing fraud.

To sum up, it seems justified to state that there is a black market for “fake” accounts and social media activity and that promotion done using this market represents an ethically gray area.

The author wants to point out that this is not to say that microworkers.com or any other platform is in itself immoral or designed to be a black market—such platforms also offer a useful venue for valid projects, like data processing at scale, acquiring subjects for survey-based or interactive online scientific research, monitoring a competitor’s public online activity, or counting objects on images -  several such campaigns were observed.

\section{Related Work}

The marketing techniques discussed in this article are related to techniques called sock puppetry and click farming. Sock puppetry means the control of many social media accounts by one person or a small group of individuals. A report on such activity was published in The New York Times \citep{Caldwell2007}. The method of control is a crucial difference in these efforts. As we will see some in Section \ref{Sec:signup} clients buy hundreds or thousands of accounts that they can use themselves. In this case there are technical possibilities to detect the puppetry, by noticing when a very high number of users are logging in from the same network location or use the browser client fingerprint \citep{Laperdrix2016}. 

But the method of control can also be an order from a client to a cohort of users to perform some activity (but the client itself never logs in to any of their accounts). In this case detection of the activity is much harder if the client takes some precautionary steps (see Section \ref{Sec:signup}). Sometimes this is called meet puppetry \citep{Cook2014} referring to the fact that it involves real freelancers.

Click farming is another related term. Click farms are actual workplaces in developing countries where a large number of employees are performing short task sometimes for as low as 1000 likes for 15 USD\citep{Arthur2013}. In current reporting these are often called troll farms \citep{Smith2018}.

The platform investigated by this article, microworkers.com differs greatly from a click farm as it is a completely distributed crowdsourcing tool, but some of the campaigns done here might be similar to those done by a click farm.

On the deceptiveness of such campaigns in comparison with traditional advertising (where the prospective customer is aware that it is being presented with an ad) is well described by \citet{DelRiego2009} and \citet{Forrest2010} in connection with then-new US Federal Trade Commission guidelines on endorsements and reviews.

This article focuses on probably way smaller market available on microworkers.com, which is, however, easily accessible for freelancers and is not specialized to gray marketing in particular.  In contrast with the services that directly offer followers and social media activity (Fiverr, SeoClerks, InterTwitter, FanMeNow, LikedSocial, SocialPresence, SocializeUk, ViralMediaBoost \citep{DeMicheli2013}), here the client has to organize its own campaign and orchestrate the freelancers on the crowdsourcing platform. This allows for creativity and innovations in the campaign methods.

\citet{Nguyen2014} explained the idea behind microworkers.com, founded in 2009, as well as reported its user count at the time of writing (presumably 2014). \citet{Howe2006} also reported on microworkers.com as a crowdsourcing platform.

According to Nguyen, the platform had over 600,000 users from 190 different countries. The aim of the microworkers.com as a project was to aid brokering crowdsourcing campaigns. As the article explained, 

\begin{quote}
\textit{In crowdsourcing platforms, there is perfect meritocracy. Especially in systems like Microworkers; age, gender, race, education, and job history do not matter, as the quality of work is all that counts; and every task is available to Users of every imaginable background. If you are capable of completing the required Microtask, you've got the job.}
\end{quote} 

The campaign templates on the landing page of the platform are great sources of inspiration for what could possibly be achieved through crowdsourcing: participating in market research, captioning documents and video, categorizing images, testing websites and applications, and so on. The fact that the majority of public campaigns visible on the platform are mostly employing controversial promotion techniques does not seem to be the result of the platform design or intentions.

\citet{Hirth2011} investigated microworkers.com in order to compare it to the much better understood Amazon Mechanical Turk \citep{Paolacci2010,Buhrmester2011}. They correctly identified a main difference between the portals: the payment mechanism. At the time, it was basically impossible to use MTurk without a US-based credit card, while the microworkers website allowed payments to be made with Moneybookers (called Skrill today). This helps explains why the author of this paper and possibly other non-US-based researchers first discovered microworkers. Works by \citet{Gardlo2012} and \citet{Crone2017} aimed to assess the usefulness of the platform for scientific purposes; and indeed, scientific projects regularly, though relatively infrequently, appeared on the platform. However, it is possible that this difference in payment methods is only one of the reasons behind the nature of the campaigns conducted on each.

Hirth et al.’s work \citep{Hirth2011} indicated that gray promotional campaigns were already existed as long ago as 2011: \enquote{Signup}, \enquote{Click or Search}, \enquote{Voting and Rating} were already featured as campaign categories; however, the payments offered were slightly higher than today.

The connection between social media and marketing was analyzed by \citet{Thackeray2008} as early as 2008. Of course this work concentrated on the legitimate social media strategies firms might embrace, such as paid search results, where the brand buys a presence in the search results. As \citet{Yang2010} explained, these are usually placed in a separate area on the results page, together with being clearly to indicate that they are paid for or they may be labeled as an ad, e.g., on Facebook (the difference between \enquote{paid} and \enquote{organic} (showing up in non-paid results) links is less emphasized in today’s search engines but remains clear). \citet{Rutz2011} demonstrated how tying paid search results to generic search terms might increase the success of a branded paid search.

The search and engage campaigns (see section \ref{Sec:search}) discussed in the present paper are different.  They don’t try to increase visibility by buying paid results. These represent the dark side of search engine-based advertising, whereby they try to directly manipulate the organic links.

It was envisaged \citep{Zhang2013} that the identification of the key influencers on social media could be crucial for effective viral marketing—but with the gray marketing techniques presented here, influence is attempted to be created directly, albeit artificially.

It was also envisioned that customer-generated content on blogs, etc. would be crucial for promotional activities—in the present paper, campaigns seeking to manufacture legitimate-looking customer content are analyzed. In other words, these campaigns, albeit unethical, are sometimes the effective counterparts of hard-to-operate social media marketing tactics, or in other words, they are controversial shortcuts to followers and likes and brand-friendly social content.

Confessore et al.'s recent work reported on in the New York Times \citep{Confessore2018} covered a very similar theme to this article, but was focused mostly on Twitter and on the activities of a company called Devumi.

\section{Aims and Methods}

\subsection{Research aims}
The primary aims of the present research were to identify the different schemes employed in microworker-based gray marketing on microworkers.com and to then categorize them, attempt to uncover how they fit in a wider strategy, estimate their limits and effectiveness, and to utilize this knowledge to provide general insights into these kinds of campaigns.

Second, the scale and typical budget of these campaigns, and their relative share of the overall activity on the microworkers.com platform were also measured and reported herein.

\subsection{Observation of campaigns}

This research project attempted to observe all campaigns posted on microworkers.com from 22 February 2016 to 21 February 2017. The site was checked several times a day during this period. In total, 7,426 campaigns were observed during the period. Each campaign was manually categorized and the aggregate numbers of categories (payments, number of tasks) were updated.

\subsection{Categorization}

The campaigns were categorized in terms of two dimensions: the related target platform (Facebook, Google Plus, SoundCloud, Twitter, etc.), and the specific activity (search and engage, like, comment, sign up, etc.). The categorization was based on the campaign title and text and proved to be quite straightforward as the names of the platforms were clearly stated in the title and represented unambiguous brand names, and as the activity to be performed was almost always explained in an itemized list in the posting. Obviously certain activities were further linked to certain platforms (like retweets can only be done through Twitter), but others, like search and engage can be done on multiple platforms.

The platform labels utilized are summarized in Table \ref{Platforms}. while the activity labels are summarized in Table \ref{Activities}.
\begin{table}[t]
\begin{center}
\begin{tabular}{| p{6.5cm} | p{6.5cm} |}
\multicolumn{2}{c}{\textbf{Platform categories}} \\
\hline
\textbf{Ali} (Alibaba, AliExpresS) & \textbf{Instagram} \\ 
\hline
\textbf{Amazon} & \textbf{Bing} \\ 
\hline
\textbf{MBS}: (microblog instant share): blogger.com, Pinterest, Digg, Tumblr, 9gag or other blogs or quick sharing platforms & \textbf{Mix}: (SoundCloud, Mixtape, datpiff) \\ 
\hline
\textbf{Browser add-on} (e.g., Chrome extensions) & \textbf{RDT} (a traffic generator site, the name of which is redacted from this article) \\ \hline
\textbf{Other}: (500px, Wordpress.com, Snapchat, Skillshare, Hotmail, LinkedIn, Coursera, Bitbucket, Snapchat, Steam, Yandex, other uncategorized) &  \textbf{Question} sites: (Yahoo Answers, Quora) \\ 
\hline
\textbf{eBay} & \textbf{Reddit} \\
\hline
\textbf{Forum} (Disqus, Warrior Forum, other forums) & \textbf{Redacted}: (not visible in description because of the rotator technique, explained below) \\
\hline
\textbf{Facebook}  & \textbf{Smartphone} (iOS and Android) \\
\hline
\textbf{Gmail} & \textbf{Twitter} \\
\hline
\textbf{Google} (search) & \textbf{Yahoo} (search) \\ 
\hline
\textbf{Google+} &  \\
\hline
\end{tabular}
\caption{Categories of platforms}
\label{Platforms}
\end{center}
\end{table}

Also, each campaign could belong to multiple activity categories, by using multiple tags.

\subsection{Limitations}

On microworkers.com, there are invite-only campaigns as well. Unfortunately, there is no information freely available on these campaigns or their share of the total number of campaigns. The nature of invite-only campaigns, for which clients apparently hire tested and trusted microworkers, could be a subject for further research.

Some campaigns might have slipped trough between two observations, meaning that their full life cycle very short (only a couple of hours). This does not appear to be typical but cannot be ruled out. Therefore, it can be said that in reality there were possibly more than 7,426 public campaigns and an additional, unknown number of invity-only ones.

As explained above, the campaigns were manually categorized by the author. This categorization, because it relied on objectively observable features of the campaigns, did not require significant subjective judgment. Therefore, in the context of intersubjectivity, it should not be a serious limitation that there was no multiple-person cross-checking performed for interpretation of the category labels. 

Finally, there are obviously other brokering platforms for such microtasks, but these are outside the scope of this investigation (in fact, some of those platforms seem to be using microworkers for recruitment purposes).

The author is confident that these limitations do not prevent the work from meeting its stated aims, as it is an explorative rather than exhaustive description of techniques and strategies.

\begin{table}[t]
\begin{center}
\begin{tabular}{| p{6.5cm} | p{6.5cm} |}
\multicolumn{2}{c}{\textbf{Activity categories}} \\
\hline
\textbf{A} Answer (a question on an answer site—see under Platforms—of a forum) & \textbf{B}	Bookmark or Pin (quick share on MBS platform—see above) \\ 
\hline
\textbf{C} Comment (Facebook, YouTube, forums, etc. comments) & \textbf{D} Data processing (counting, summarizing information) \\
\hline
\textbf{E} Engage (usually unspecified web activity, e.g., “use the website for a while” or social media activity that cannot be categorized in the other labels) & \textbf{F} Follow (various social media, mostly Twitter) \\
\hline
\textbf{H} Share (using the share function on various social media) & \textbf{I} Install (install applications to a smartphone or computer) \\
\hline
\textbf{K} Link (add a given link to a comment, question answer, share) & \textbf{L} Like/Upvote (depending on platform: YouTube thumbs up, Facebook like, Reddit upvote, Google+ +1) \\
\hline
\textbf{M} Upload or Download (Upload: YouTube videos, Download: various files) & \textbf{N} Connect/Friend (Depends on social media platform, e.g., on Facebook, connects to become friends) \\
\hline
\textbf{O} Other (anything that could not be categorized otherwise) & \textbf{P} Write/Post/Blogpost (create and post written content, similar to comment but usually longer) \\
\hline
\textbf{Q} Ask a question (on an answer site—see under Platforms) & \textbf{R} Research participation/Survey \\
\hline
\textbf{S} Search/Search and Click/Search and Visit (use the search function: Google, Yahoo, Bing, Facebook search, others) & \textbf{T} Tweet or Retweet \\
\hline
\textbf{U} Signup (YouTube channel, mailing list, portal, etc.). Sometimes this involves handing over the login credentials & \textbf{V} Vote (vote on a given entrant, various voting platforms) \\
\hline
\textbf{W} Watch (usually YouTube videos) & \textbf{Y} Captcha (solve capthas) \\
\hline
\textbf{Z} Test (software or website testing) & \\
\hline
\end{tabular}
\caption{Categories of activities}
\label{Activities}
\end{center}
\end{table}

\subsection{Anonymization}

Generally, all data presented in this paper (mostly microtask descriptions) is anonymized. Most of the actual URLs, person and company names, and other named entities are replaced with the string $<REDACTED(...)>$. When there are several URLs or names within one example though, they are replaced with their own unique label so that they are not mixed up. Other than this modification, the job descriptions are copied herein verbatim.

Obviously, some basic URLs like Facebook.com or fakenamegenerator.com, for example, are kept because they are reported only for uncovering the campaign method but not its content, and also because the job descriptions would not be as understandable without them.

The reason for the redaction is that the persons, websites, and Facebook accounts mentioned in these task descriptions may be unwilling targets of a campaign. It is also probable that the customers of promotion campaigns are often not aware or may even have been misled about the methods employed on their behalf.

\section{Results}

Based on the observations, the following summaries were created.

Table \ref{ActBudg} contains budget summaries by activity. The columns represent the activity code, number of campaigns, number of tasks, and net budget (without the 10\% fee). The table is ordered by the descending number of campaigns.

\begin{table}[h]
\begin{center}
\begin{tabular}{| l | l | l | l |}
\hline
\textbf{Activity} & \textbf{\# campaigns} & \textbf{\# tasks} & \textbf{total budget} \\
\hline
L & 1,303 & 207,811 & \$ 22,757.32 \\
\hline
P &	1,293 & 116,682 & \$ 26,796.95 \\
\hline
S &	1,229 & 577,444 & \$ 46,802.91 \\
\hline
U & 1,210 & 354,180 & \$ 40,344.71 \\
\hline
C & 733 & 227,756 & \$ 27,926.15 \\
\hline
E & 495 & 318,387 & \$ 23,310.10 \\
\hline
I & 361 & 16,934 & \$ 8,375.01 \\
\hline
H & 357 & 26,305 & \$ 7,547.54 \\
\hline
Z & 352 & 122,974 & \$ 9,681.65 \\
\hline
N & 203 & 26,808 & \$ 3,419.92 \\
\hline
B &	196 & 40,045 & \$ 4,422.94 \\
\hline
W & 147 & 38,756 & \$ 4,307.35 \\
\hline
D & 138 & 50,886 & \$ 6,307.99 \\
\hline
F & 79 & 12,649 & \$ 1,520.29 \\
\hline
V & 68 & 32,832 & \$ 4,026.03 \\
\hline
T & 31 & 2,071 & \$ 409.87 \\
\hline
R & 29 & 5,191 & \$ 2,037.10 \\
\hline
O & 24 & 7,219 & \$ 1,240.66 \\
\hline
A & 14 & 519 & \$ 78.90 \\
\hline
M & 12 & 847 & \$ 203.45 \\
\hline
K & 10 & 474 & \$ 292.02 \\
\hline
Q & 3 & 150 & \$ 24.60 \\
\hline
Y & 1 & 1,026 & \$ 61.56 \\
\hline
\end{tabular}
\caption{Budget Summary by Activities}
\label{ActBudg}
\end{center}
\end{table}

If we take out the category data processing, research participation, captcha solving, testing, installing and \enquote{other}, the remaining activities are purely for promotional purposes. This leaves 1,665,138 tasks, or 89.7\% of the whole. It should be added that many of the install tasks appear to be promotional (see section \ref{Smartphone}). Counting these in would make the figure even higher. However, since for many such campaigns this aspect is impossible to tell, they are left out.

A similar summary for the platforms is given in table \ref{PlatBudg}. The first column is the platform name, the rest is the same as before. The table is ordered by the descending number of campaigns.

\begin{table}[h]
\begin{center}
\begin{tabular}{| l | l | l | l |}
\hline
\textbf{Platform} & \textbf{\# campaigns} & \textbf{\# tasks} & \textbf{\# total budget} \\
\hline
Smartphone & 1102 & 231,892 & \$ 27,702.07 \\
\hline
Google+ & 823 & 57,203 & \$ 18,960.50 \\
\hline
Other & 797 & 349,937 & \$ 31,333.06 \\
\hline
Twitter & 747 & 83,731 & \$ 14,047.21 \\
\hline
Facebook & 666 & 115,967 & \$ 15,657.86 \\
\hline
YouTube & 616 & 248,914 & \$ 29,322.28 \\
\hline
Reddit & 539 & 77,104 & \$ 5,434.15 \\
\hline
Redacted & 501 & 132,401 & \$ 13,397.94 \\
\hline
Google & 330 & 261,923 & \$ 18,672.37 \\
\hline
RDT & 296 & 80,844 & \$ 8,034.94 \\
\hline
MBS & 218 & 38,859 & \$ 5,497.36 \\
\hline
Instagram & 172 & 19,421 & \$ 2,072.35 \\
\hline
Amazon & 156 & 55,760 & \$ 6,934.85 \\
\hline
Mix & 144 & 11,803 & \$ 1,458.55 \\
\hline
Question site & 126 & 16,190 & \$ 2,447.59 \\
\hline
Forum & 84 & 7,558 & \$ 964.15 \\
\hline
Gmail & 50 & 10,275 & \$ 2,908.35 \\
\hline
eBay & 22 & 6,014 & \$ 570.52 \\
\hline
Yahoo & 21 & 45,686 & \$ 2,355.13 \\
\hline
Ali & 9 & 3,945 & \$ 551.91 \\
\hline
Bing & 4 & 675 & \$ 69.00 \\
\hline
Browser add-on & 3 & 214 & \$ 73.96 \\
\hline
\end{tabular}
\caption{Budget Summary by Platforms}
\label{PlatBudg}
\end{center}
\end{table}
From these values, the average payment for tasks related to certain activities and platforms could be calculated. Here is the distribution of the payments (not equal ranges):

\begin{table}[h]
\begin{center}
\begin{tabular}{| l | l |}
\hline
\textbf{task payment} & \textbf{total number of tasks} \\
\hline
\$0 & 99,999 \\
\hline
\$0.05–\$0.1 & 1,059,172 \\
\hline
\$0.11–\$0.2 & 581,146 \\
\hline
\$0.21–\$0.3 & 63,904 \\
\hline
\$0.31–\$0.5 & 35,893 \\
\hline
\$0.51–\$1.0 & 15,671 \\
\hline
\$1.1–\$3.0 & 531 \\
\hline
\end{tabular}
\caption{Number of tasks by payment range}
\label{Payment}
\end{center}
\end{table}

The lowest paid wage was \$0, and this was incidentally the biggest campaign with 99,999 tasks. The task was a simple visit to a link. The client promised future tasks for those who completed the task. More detail is provided on this task in section \ref{Sec:signup}.

The highest paid wage was \$3 for a task, but as can be seen from the above table, there were only 531 jobs offering between \$1.1 and \$3, while there were over a million jobs offering between 5 and 10 cents, making the higher paid tasks very rare indeed.

\section{Campaign types and their analysis}

\subsection{Voting}

Participation in voting (V) is a recurring activity on microworksers.com, with 68 campaigns posted featuring an aggregated 32,832 votes purchased. The top two voting campaigns seemed to be promoting a product and a sports team (2,130 and 2,000 tasks), the third was a giveaway voting for an expensive trip for couples, where the entrants were supposed to vote on the videos they uploaded about themselves. One entrant purchased 1,703 votes (the wording reveals that they bought the vote for themselves personally) for \$0.12 each:

\begin{lstlisting}[caption=Buying votes, label=votes]
Title: Video: vote
Payment: 0.12
Number of workers accepted: 1703

1. Go to <REDACTED URL>
Give the video a VOTE by clicking on the heart below the video
Required proof that task was finished?
1. Tell me how many votes I had after you've voted
\end{lstlisting}

Unfortunately while this can be seen as being clearly unethical, for a little more than \$ 200 it could have been economically viable if the entrant won the vote. And it is even plausible that a local vote could have been won with just 1,703 extra votes bought.

Other votes were for titles like best auto repair shop, best bakery or \enquote{tradie of the year in Australia}. There was also a census on how many New York City residents wanted to go on a date with a certain model. There were votes on temple photography, the best fintech firms, music mixes, the best female vocalists, several contests about the ranking of attractive persons, a vote on XXL Magazine, a vote on the best local charter flight provider, and so on. Some of these were clearly promoting a product or a performer or artist; others seemed to be clearly what we will call vanity-promotion.

The purchased votes ranged from dozens to about 1,000 at a time. It is hard to assess the overall efficiency of such campaigns, but it can certainly be said that for local contests, where the maximum number of people voting is expected to be measured in hundreds or thousands, it is very easy to rig contests this way, as it would cost only a few dollars. 

\subsection{Search and engage tasks}
\label{Sec:search}

The common feature of these kinds of microtasks seems to be an attempt to manipulate the search results in search engines like Google, Yahoo, Bing or the search feature of Facebook. In most cases, a certain item is promoted, but in some rare cases, the goal actually seems to be to push unwanted result items back in the result list.

These campaigns appear to assume that search engines learn: if for given search term, a high number of users click on a particular result item, then that result item must be a good result for the search and therefore it will be listed early in the results listing. While the actual algorithms search engines use are proprietary and unpublished, it is known how they work in theory \citep{Buettcher2016}. It is thus plausible that they can be tricked in this way to a certain extent. We know that user behavior is taken into account in Google, for instance, as \citep{Clark2015} reported that Google’s novel AI solution, BrainRank, was the third most important factor (the technical term is \enquote{signal}) when ranking pages. We also know that it learns from user behavior, hence it is plausible these systems can be tricked through paid user behavior simulating genuine interest.

A search and engage campaign therefore hires a large number of microworkers for searching the given terms and clicking on the promoted item in the search results.

An example of this type of campaign is given below:

\begin{lstlisting}[caption=Google search, label=googlesearch]
Title: Google search
Payment: 0.08
Number of workers accepted: 1600

Job description:
1. Open up Google.com (please use US version)
2. In Google, please search for this phrase: <REDACTED PERSON NAME>
3. Please click on any of the red boxed links you see in the attached file <the file is a Google result list screenshot, red rectangles designate what result items need to be promoted>
4. Stay on page for 1 minute
(...)
\end{lstlisting}

The top ten search and engage campaigns have the number of tasks offered as between 2,941 and 6,770. However, many of these campaigns seem to be part of the same project, making the biggest projects around 10,000–20,000 tasks.

An interesting tendency is that in many of these promotion campaigns, it seems that there is no marketed product involved, rather it is individuals concerned with their online persona who are the payers, in what is really another example of vanity-promotion.

The biggest search and engage project, with above 10,000 tasks, involved the promotion of a USA business executive’s Wikipedia entry, whose name happens to be the same as a famous USA American football player and also a former USA congressman. The project must have been a success as currently the promoted page comes out top in a Google search when searching for that name. Naturally, it is impossible to establish the causal relation between the campaign and the current ranking with any certainty, especially this long after the campaign.

\subsection{Social media activity}

Paid social media activity involves task like creating Pinterest Pins, upvoting in Reddit, YouTube, or Google+, liking in Facebook, using Digg, Twitter, or Instagram, commenting on forums, upvoting on SoundCloud, Mixcloud or Datpiff, and so on. In terms of activity codes, this section covers B, C, F, H, L, N, T, W.

Campaign example~\ref{facelikes} is in this category. The campaigns are usually straightforward and easy to do, therefore the payment is usually very low. Clicking on like, upvote, etc. are the lowest paid tasks. For instance, the 77,104 Reddit upvotes purchased during the 365 days study period cost less than \$ 5,500 in total (see Table \ref{PlatBudg}). The highest paying jobs in this category were those that require writing content that meets a set specification, e.g.:

\begin{lstlisting}[caption=Youtube comments, label=youcomment]
Title: YouTube: Comment 3x (1-3)
Payment: 0.30
Number of workers accepted: 90

Job description:
1. Go to www.youtube.com/channel/<REDACTED>.
2. Post a positive relevant comments on the videos found in first three links
Important: Comment must be at least 10-15 words long and cannot be generic and must include the following words <REDACTED Person name> and the word "Florida".
3. Stay on each video page for 1 minute

Required proof that task was finished?
1. YouTube display name
2. Copy of the comments you've posted
3. URL to YouTube videos where comments were posted
\end{lstlisting}

However, in other cases the freelancer is asked to copy-paste the comment content, and the job then pays less:

\begin{lstlisting}[caption=Youtube comments - the copy-paste method, label=youcommentcc]
Title: YouTube: Comment 3x (<REDACTED>)
Payment: 0.12
Number of workers accepted: 440

Job description:
1. Go to the instruction page: <REDACTED URL>
2. Search Youtube.com for the key phrase
3. Copy-paste the supplied comments onto relevant video
4. Repeat steps 2 and 3 for 2 more videos (3 total)

Required proof that task was finished?
1. Your YouTube name
2. The search phrases you used
3. URLs of the 3 videos you commented on
\end{lstlisting}

For commenting, the most prominent platforms are YouTube (364 campaigns; 197,735 tasks, some paying for three comments), Instagram (101; 4,670), questions sites (81; 10,282), Facebook (30, 2,485), and a long tail of other forums (see some under Platforms; 118 campaigns).

Following a given account is done on Google+ (317 campaigns; 21,228 tasks) Instagram (54; 8,960), Twitter (20; 2325), Quora and Yahoo Answers (4; 466), and Google+ (1; 898). In must be noted that for Twitter, the clients often require the presence of some features from the freelancer, e.g.:

\begin{lstlisting}[caption=Twitter account quality rules, label=twittercheck]
(...)To do this task, you need to have a Twitter account that meets the following requirements:- At least 100 followers - Your follower count needs to be at least double your following count (meaning - if you are following 100 users, you need to have at least 100*2 =200 followers) - The majority of the 20 most recent tweets are in English - At least 8 out of 20 most recent tweets have no links, are Not Retweets, and sound natural and interesting.(...)
\end{lstlisting}

This is obviously requested in an effort to imitate a real Twitter user and to not seem like a newly created one. These tasks pay bonuses too, meaning that the pay can reach as high as \$ 0.25.

Posting (P) and Tweeting (T) were grouped together for being very similar. P and T is most prevalent on Twitter (716 campaigns; 80,211 tasks) and Google+ (485; 28,819).

Connecting as a friend is mostly done on Facebook (13 campaigns; 9,190 tasks).

Liking/Upvoting (L) is an activity performed on Reddit (539 campaigns; 77,104 tasks), Facebook (485; 71,200), Mixcloud and SoundCloud and Datpiff (143; 11,773), YouTube (63; 23,993), Instagram and Google+ (both 13 campaigns, 5,366 and 5,401 tasks, respectively), and on some other platforms (49).

The 10 biggest like/upvote campaigns ranged between 1,830 and 7,417 offered tasks (here, Facebook, Reddit, Instagram, Google+ campaigns were all in the top 10). The content of these top campaigns unfortunately were redacted using the rotator technique (see later in the article), but some of the remaining involved promoting persons not noted on Wikipedia (vanity-promoting); and some niche products. Size seems to be a limitation again, just like with search and engage campaigns: for celebrities with hundreds of thousands of followers, even as much as 7 thousand new likes hardly seem to matter, and the promotion technique does not seem to scale up to higher numbers.

This limitation might be not there in the case of comments (C) campaigns. The biggest comments campaign was conducted on YouTube, very similar to example~\ref{youcommentcc}, and it involved 21,140 tasks, three comments each, yielding more than 63,000 paid comments. The tenth-biggest involved 2,425 tasks, again three comments each. While a similar amount of likes would still represent just a fraction when it comes to comparing it to the likes received for the most popular YouTube videos, Facebook accounts, etc. For comments, the case is different, because only a small fraction of readers/visitors make comments. A cursory, non-representative investigation of the many YouTube videos reveals that it is very hard to find videos that have less than 20x viewers than comments. Thinking the other way around, 63,000 tendentious comments would suggest representing well over 1.2 million viewers, hence distorting the perception of actual viewers on what other’s opinions are in relation to the topic.

However, in local communities with a smaller overall size, it seems that even small (L) campaigns can make a difference among the competition. Consider this example related to warriorforum.com (its self-description is: \enquote{The world's largest Internet Marketing Community and Marketplace.})

\begin{lstlisting}[caption=Comments on a small forum, label=warriorforum]
Title: Warriorforum Post: Comment
Payment: 0.10
Number of workers accepted: 30

Job description:
Must have a Warriorforum account at least 2 months old, or a very active account if you joined recently.
1. Go to: www.warriorforum.com/<REDACTED>
2. Post a comment/testimony relating to the post
Must be original/no copying others post
Tip on what you can post:
I learned a lot from the webinar,
I just join and I am excited, cannot wait to start working with <REDACTED>,
this is an awesome program, etc.

Required proof that task was finished?
1. Name used to leave comment
\end{lstlisting}

A cursory look at Warrior Forum reveals that the typical number of upvotes is around 10 and the number of comments (reply-s) is similar. The job offer above for 30 comments in example~\ref{warriorforum} would thus propel the entry among the top, while costing a mere \$ 3.3 for the client.

\subsection{Smartphone apps}
\label{Smartphone}

A distinct area of paid activity is installing smartphone applications, using/testing and rating them. There were 361 such campaigns with 16,934 tasks worth \$ 8,375.01. A typical campaign looks like this:

\begin{lstlisting}[caption=Smartphone app "testing", label=smartphone]
Title: Android App Testing (<REDACTED>): Download+Install+Honest Review
Payment: 0.50
Number of workers accepted: 370

Job description:
1. Install the app below: play.google.com/store/apps/<REDACTED>
2. Download the app
3. Open the app for 30 seconds and test
4. Rate the app
Optional: Leave 3, 4, or 5 stars
5. Write an honest review in the Microworkers proof box only

Required proof that task was finished?
1. Your Google username
2. Paste your review
\end{lstlisting}

We can interpret this campaign in several ways. The charitable interpretation would be that this is an honest test for the software. Even though the app is only required to run for 30 seconds, the freelancers would open it on dozens of different Android devices, with different capabilities, screens, API levels, etc. Thirty seconds is enough to run some self-assessment and report back to a server. This test could thus have some value from a Software Engineering perspective and may be a valid assignment.

However, this kind of test is surely already long overdue in the case of a  published application. Issues around crashing apps and unwanted startup behavior should have been resolved way before then. The author speculates that these kinds of campaigns are instead promotional. The clients are usually careful not to explicitly order five stars or positive reviews (albeit there are counter-examples of such). Yet, in such a campaign, an initial, visible user base is created for the app. Again, we can assume that this is more useful in niches than in competition with mainstream applications, as the leading apps have millions of users already.

It must be mentioned that this kind of microtask carries security risks for the freelancer and for the general public. The fact that the freelancer installs apps for a fraction of a dollar presents an obvious opportunity for breaching their smartphones. Although the current number of tasks in these campaigns does not seem to be high enough, in theory it would be possible to create a zombie network for DOS attacks or similar purposes.

\subsection{Signup}
\label{Sec:signup}

A very common type of campaign is the signup (U) campaign. These campaigns involve creating an account meeting some client-specified requirements. In many cases, the microworkers will be required to hand over the account credentials. Example~\ref{youtubeacc} below was one of the biggest signup and account handover campaigns observed.

\begin{lstlisting}[caption=Youtube Account Creation, label=youtubeacc]
Title: YouTube: Create an Account
Payment: 0.10
Number of workers accepted: 2290

Job description:
YouTube: Create an Account
1. Go to www.youtube.com
2. Create a new account
3. Verify your email
4. Login to activate the account

Required proof that task was finished?
1. YouTube username or email
2. YouTube password
Your task will NOT be rated satisfied if your YouTube account requests phone verification.
\end{lstlisting}

The client in this case has acquired 2,290 YouTube accounts for a mere gross \$ 251.9. Example~\ref{gmailacc} from the introduction is a similar case, but for Gmail. The dangers posed by such campaigns are obvious. Besides promoting products, ideas and agendas, a cohort of 2,300 YouTube users can disrupt any smaller community on the platform and the use of fake accounts could facilitate the account owner to commit fraud or otherwise abuse the system.

What makes these mass account acquisitions very dangerous is that they are not easy to detect. Methods that are able to detect fake accounts typically only work if they are all created by the same person \citep{Xiao2015}, but cannot be expected to work in this case as these accounts are created by real people. A landmark study by \citet{Gurajala2015} involving the analysis of 62 million Twitter public user profiles relied on statistics about update frequency, reused profile pictures, and account creation days. Unfortunately, these factors can all be made to look genuine; for instance, the freelancers can be instructed to use profile pictures that are not reused; or possibly profile pictures themselves are acquired via microworkers (see example~\ref{photos}), and the creation times can be spread out with the help of \enquote{throttling}—a feature of the platform that allows only a certain number of tasks to be completed in a unit of time. Sometimes clients give instructions that enable the detection of such accounts, e.g., by requiring the freelancers to use the very same password. Also, it is probable that after handing over an account, the geolocation of the usage of that account is changed permanently, and so never again reflects the country of creation, which could be a factor in detection.

Not all signup campaigns seem to require account handover. For instance, the top four signup campaigns in terms of task numbers required a signup to two different website traffic providers and a polling site; involving 21,388, 19,952, 10,888, and 7,648 individual signups. Related to these campaigns was the biggest (99,999 workers) and cheapest (paying \$0.0) campaign observed, categorized as testing (Z), as technically it was a website spellcheck; its details are given in the following example:
\begin{lstlisting}[caption=A recruitment campaign involving a small test to pass, label=recruit]
Title: Qualification Test: Find the Misspelled Word
Payment: 0.0
Number of workers accepted: 99 999

This is a qualification for a future website test which will pay $7.50 for about 11 minutes of work. This qualification test is to find workers with a great eye for detail.
1. Go to REDACTED URL
2. Find the misspelled word.
Hint: it is near the bottom
Required proof that task was finished?
1. The wrongly spelled word in its wrongly spelled form
\end{lstlisting}

All five sites (the aforementioned four traffic providers and the one in Example~\ref{recruit}) were categorized as \enquote{Other} on the platform, and were not commonly featured. The scale of these campaigns explain why the category Other is so prominent in the platform aggregation in Table~\ref{PlatBudg}. Also, all the sites are basically recruiting microworkers for their own platform. The nature of tasks to be done there seems to be are traffic generation (visiting sites), participation in paid market research by answering surveys, etc. 

Among the next five in the top 10 signup campaigns (places 6-10 ranked by the number of tasks on offer) was example~\ref{youtubeacc}, another account creation and handover involving 2,190 accounts to a site redacted with the rotator technique, plus three jobs requiring the signup of several thousands of users to various sites for unknown reasons.

\subsection{Other interesting campaigns}

This section covers several interesting campaigns that cannot easily be categorized in the other categories, many of them one-of-a-kind campaigns, and some of them seem rather strange and unexplained.

There was one campaign that requires the users to solve captchas. This is obviously to bypass a captcha-protected signup page. We can hypothesize that this is part of a human-in-the-loop automated account creator system. 

There were several research campaigns observed. These are transparent and benign: the university or the research group is clearly present, there is usually a document attached as a brief for the research. The topic seems to be social psychology or web usability and ergonomy. The users are asked their gender and then made to do face expression recognition; evaluate risks; try out different webpage workflows, etc.

There is one observed snapchat promoter recruitment campaign (30 tasks x USD 0.5), see the following:

\begin{lstlisting}[caption=Snapchat recruitment, label=snapchat]
Objectives: I'm looking for cute girls who snap for marketing promotions (bonus possible).
Important: You must actually snap video and not just upload pictures from fake profiles.
1. Provide your Snapchat Username for proof (I will add you as a friend)
\end{lstlisting}

Some campaigns seem to be building stock photos, like example~\ref{photos} below (1000 x \$ 0.11 ). Another project required photos of windshields. Yet another project asked for a selfie of the freelancer, and the consent to use it, but only from those who had no beard.

\begin{lstlisting}[caption=Acquiring photos, label=photos]
Important: You must agree to allow us to use your photo for promotional purposes in order to complete this task.
1. Take a well lit, clear photo of an office. Important: Make sure no people are in the photo.
Notes:- I need a clear photo of the office showing computers, desks, etc.- Photo should not be a photo from the Internet, we search for all photos on the Internet before we approve
\end{lstlisting}

Finally, for the following campaign there is just no explanation:

\begin{lstlisting}[caption=An unexplained campaign, label=anthem]
(30 x $1.75) Write 12 lines of lyrics for an Anthem based on your own individual traditions and struggles (Make it relevant to your life today) (...)
\end{lstlisting}

\section{Techniques employed in campaigns}

As explained in the Limitations section, there are a number of invite-only campaigns on the site, called \enquote{hire groups}. These allow a client to select the freelancers, as contrasted to public campaigns that are open to anyone to participate. Also, this allows a per-employee task customization by providing a spreadsheet of input variable values. While also being feature rich, hire group campaigns are usually hidden from the public view.

Rotators are another way of per-employee customization and also allows hiding the content of the campaign from public display.

\begin{lstlisting}[caption=The rotator technique, label=rotator]
Title: Forum: Sign up + Post + Screenshot
Payment: 0.14
Number of workers accepted: 96
1. Go to this link: bit.ly/<REDACTED URL ENDING1>
2. Search for blogs from this search link
3. Find blog, website or forum you can post a comment on
4. Go to this link: bit.ly/<REDACTED URL ENDING2> and then copy comment from this page and post this comment in the website blog or forum

Required proof that task was finished?
1. Your Forum Username
2. URL of the comment
3. Screenshot of posting
\end{lstlisting}

This technique allows the employer to customize the task per-employer without a hire group and to remove the instructions after the campaign is done without leaving a trace. Except for those freelancers who participated in the campaign, there is no way of knowing what sites, search keywords, or comments were involved in the job. The category \enquote{Redacted} among the platforms refer to this technique and not to data anonymization employed by the author in the examples in this article. Of course, there is no way of knowing if the client’s intention was just to rely on task customization, or to hide the campaign content, or both.

As explained at the Smartphone apps section, there might be ways for dressing up promotion campaigns as testing campaigns, by asking a couple of hundred users to install the app and then leave it there. Also, there might be search and engage campaigns masquerading as data collection and competition monitoring. In the case of some Amazon- and eBay-related campaigns, the freelancers are directed to search for different products, then to select from a given set of results, and then to collect prices, data, specifications, and to finally submit these as job proof. What makes these suspicious is that for an honest information campaign, it seems to be overly redundant to collect the same information many hundreds of times by many hundreds of microworkers. In reality, the point of these campaigns could be to make the microworker search and engage and then to spend time on the visited page while counting reviews and collecting information (the algorithm of a search engine might take the duration spent on a result page into account when adjusting itself), and then the accomplishment of the job can be conveniently verified by the client by looking at the collected data. Of course, these are just hypotheses for which there is no way to verify them.

Figure \ref{Fig1} summarizes the promotional methods observed, together with the supporting techniques featured in various gray promotional campaigns:
\begin{figure}[!bt]
\centering
\includegraphics[width=1\hsize]{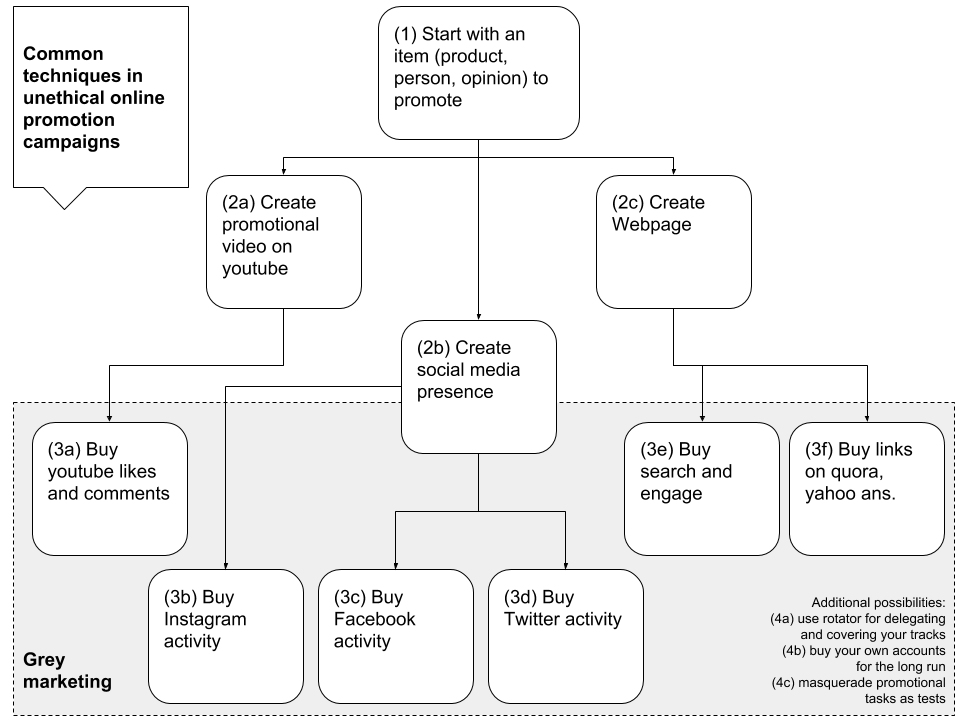}
\caption{Elements of unethical online promotion campaigns.}
\label{Fig1}
\end{figure}

\section{Conclusion}

This article provides insights into the black market of likes, upvotes, comments, retweets, votes in contests, and search engine manipulation. The subject of the investigation was microworkers.com, which is not a black market itself per se, but it has light regulation of its campaigns and so can be used by clients to participate in black-market activities. Also, it clearly is only one of several venues for running such campaigns. \citet{DeMicheli2013} have investigated several other players on the market (Fiverr, SeoClerks, InterTwitter, FanMeNow, LikedSocial, SocialPresence, SocializeUk, ViralMediaBoost) and they have found that the market size is probably several millions of dollars, making the share microworkers.com a tiny fraction. Other sites even recruit on microworkers.com for similar microtasks. However, thanks to the fact that on microworkers.com the client has to orchestrate the campaigns itself, we can get an insight how the other players in the market, that sell complete like and follower packages, might be operating.

The nature of the microworkers.com campaigns was explained in the sections above. About their efficiency, we concluded that it probably varies. The main limitation is that it seems to be hard to purchase more than some tens of thousands of items. As explained in the section on Social media activity covering likes/upvotes (\textbf{L}), these numbers do not make a big difference when it comes to widely discussed political topics or celebrities, as in this area millions of \textbf{L} items are not uncommon. However, in smaller communities, with normally dozens or hundreds of \textbf{L} items, they can make a huge difference. This is the context in which the effects of a total of 207,811 \textbf{L} tasks can be assessed. For instance, Reddit, on which 77,104 upvotes were purchased, is a platform where a couple of hundred or thousand purchased upvotes can go quite far, especially in thematic sub-Reddits. It might be noted that a similar number of downvotes would be much more significant as there are normally much less of these items—but no downvoting/dislike campaigns were observed.

For comments, we have to assume that the big observed campaigns, reaching 60,000 YouTube comments must be effective as these are quite high numbers when it comes to comments. It is of course unclear what the overall effect of tens of thousands of comments is on the thinking of the targeted audience. But it is enough to provide an apparent majority on almost any platform.

For online voting and contests, it seems that all kinds except the biggest contests can be rigged by microworker campaigns.

There are two areas where the efficiency is especially hard to assess: search and engage and app testing. For search and engage, over half a million tasks were observed and we must assume that the efficiency of these jobs really depends on the popularity of the topic in question. Also, the effect of these campaigns is really hard to track. In a similar way, a hypothesis was provided on how app developers on Android or iPhone might by trying to build an initial user base of a couple of hundred installs. The most popular applications have tens of millions of user and even their alternatives often have tens or hundreds of thousands (this also indicates just how hard the entry must be to that market). A better understanding of the app market places would be necessary to understand the significance of a couple of hundred individual users.

Finally, the knowledge that several thousand YouTube, Gmail, Snapchat and other accounts have been created and their usernames and passwords handed over during the period of observation is very troubling. Those accounts might be effectively used for large scale gray promotion campaigns and could also pose a security threat at the same time.

Future work could involve participatory research as a freelancer on this or other platforms to reveal the experience of a microworker as well as to discover more about the invite-only/hire-only campaigns. In cases of comment copy-pasting, the source and nature of the comments could also be learned. Another area could be investigating the logic and goals behind the traffic generator sites and unions that similarly to microworkers’ sites rely on the completion of menial tasks.

\bibliography{OpMicro}

\end{document}